\documentclass[11pt,a4paper]{article}    

\usepackage{a4}

\usepackage[footnotesize]{caption2}
\usepackage[footnotesize]{subfigure} 
\usepackage[bookmarks]{hyperref}

\newif\ifpdf
	\ifx\pdfoutput\undefined
	\pdffalse 				
	\else
	\pdfoutput=1 			
	\pdftrue
	\fi

	\ifpdf
	\usepackage[pdftex]{graphicx}
	\else
	\usepackage{graphicx}
	\fi
 \ifpdf
	\DeclareGraphicsExtensions{.pdf, .jpg, .tif}
	\else
	\DeclareGraphicsExtensions{.eps, .jpg}
	\fi

\begin{document}

\title{Small and Large Scale Fluctuations in Atmospheric Wind Speeds}
\author{F. B\"ottcher, St. Barth \& J. Peinke \\ 
ForWind -- Center for Wind Energy Research\\ Institute of Physics, 
C.v.O. University of Oldenburg, \\
D 26111 Oldenburg, Germany} 
\date{\today} 
\maketitle

\begin{abstract}

\noindent Atmospheric wind speeds and  their fluctuations at different locations (onshore and offshore) are examined. One of the most striking features is the marked intermittency of probability density functions (PDF) of velocity differences -- no matter what location is considered. The shape of these PDFs is found to be robust over a wide range of scales which seems to contradict the mathematical concept of stability where a Gaussian distribution should be the limiting one.\\
Motivated by the instationarity of atmospheric winds it is shown that the intermittent distributions can be understood as a superposition of different subsets of isotropic turbulence. Thus we suggest a simple stochastic model to reproduce the measured statistics of wind speed fluctuations. 
\end{abstract}

\section{Introduction}  

Atmospheric wind may be seen as a prime example of a turbulent velocity field with very high Reynolds numbers of about $Re\approx10^{8}$ \cite{burton}. Reynolds numbers as large as this prevent analytical calculations and direct numerical simulations. Therefore the flow has to be described in a statistical way. For the estimation of extreme loads as well as for risk estimations the statistics of velocity fluctuations $u(t)$ and velocity differences should be known. It has been shown that these statistics obey non-Gaussian, intermittent distributions (e.g. \cite{riso}, \cite{kantz}) that directly correspond to an increased number of wind gusts \cite{us}.\\
Nevertheless, for most technical and meteorological problems fluctuations as well as fluctuation differences are assumed to obey Gaussian statistics. Therefore simulations of atmospheric velocities are often based on Gaussian processes \cite{bier}. \\
Fluctuation differences are commonly measured using {\it velocity increments}:
\begin{eqnarray}
	u_{\tau}(t) := u(t+\tau)-u(t) \;\;\; .
	\label{inc}
\end{eqnarray}
Large increment values can be identified as wind gusts as long as the time step $\tau$ is rather small. The demand for a small $\tau$-value (typically less than a minute) is due to the fact that gusts are related to large velocity rises during short times. Rises that occur over time steps of  several hours  -- for instance -- are not called wind gusts but large scale variations.\\ 
The difficulty to fix a suitable time scale mirrors the fact that atmospheric winds exhibit variations on any time scale -- in principle ranging from seconds (and less) up to centuries. For most practical applications, such as engineering and  meteorology one mainly distinguishes between large scale variations such as diurnal, weekly and seasonal changes and variations on small scales often referred to as {\it atmospheric turbulence} or {\it gustiness} \cite {burton}. The existence of a {\it mesoscale gap} as proposed by \cite{vander} which divides small (micro) and large (macro) scales in a more rigorous way has strongly been debated in recent years (e.g. \cite{lode1}, \cite{bush}).\\

\noindent In this paper we focus on the scale dependent statistics of atmospheric increments and compare them to that of homogenous, isotropic and stationary turbulence\footnote{For simplicity we will use the term 'isotropic turbulence' instead of homogenous, isotropic and stationary turbulence.} as realized in laboratory experiments. For isotropic turbulence the statistical moments of increments, the so-called structure functions have been intensively studied \cite{frisch}. Their functional dependence on the scale $\tau$ is described by a variety of multifractal models. Besides the analysis of moments, probability density functions (PDFs) are often considered. These show a transition from Gaussian distributions to intermittent (heavy-tailed) ones as scale decreases. Unfortunately the analysis of moments as well as that of probability density functions $p(u_{\tau})$ is more or less restricted to the {\it inertial range} -- the range of scales larger than the Taylor scale $\Theta$ (where dissipation effects become significant) and smaller than the integral scale\footnote{Normally $\Theta$ and $T$ denote length scales. For constant mean velocities and applying Taylor's hypothesis of frozen turbulence length- can be defined as time-scales as well. Here we will proceed with corresponding time scales.} $T$.\\
The challenge is to describe and to explain the measured fat-tailed distributions and the corresponding non-convergence to Gaussian statistics. Large increment values in the tails directly correspond to an increased probability\footnote{The probability to observe an increment $u_{\tau}$ $\epsilon$ $[u_{\tau}, u_{\tau}+du_{\tau}]$ is just given by $p(u_{\tau})du_{\tau}$.} (risk) to observe large and very large events (gusts).  As pointed out in \cite{adp} the probability to observe large events -- e.g. events twice as large as a reference -- can become negligible for a Gaussian while for heavy-tailed distributions there is still a significant probability to observe it.\\
The atmospheric PDFs -- we examine here -- differ from those of turbulent laboratory flows where -- with decreasing scale -- a change of shape of the PDFs is observed (e.g. \cite{castaing}). For large scales the distributions are Gaussian while for small scales they are found to be intermittent. The atmospheric PDFs however change their shape only for the smallest and then stay intermittent for a broad range of scales. Such a constant shape for larger and larger scales is expected only for stable distributions such as Gaussian ones or the L\'evy stable laws \cite{sornette}. Although the decay of the tails indicates that distributions should approach Gaussian ones (as for isotropic turbulence) they show a rather robust exponential-like decay. This point will be clarified in chapter \ref{prob}.\\
In chapter \ref{sca} and \ref{prob} it will be shown that atmospheric increments behave quite similar to those of isotropic turbulence for small scales but differ significantly for large ones. We therefore introduce a model -- chapter \ref{super} -- that interprets atmospheric increment statistics as a large scale mixture of subsets of isotropic statistics. When mixing is weak the same statistics as for isotropic turbulence is recovered while for strong mixing robust intermittency is obtained.  
In chapter \ref{conc} the results are briefly discussed.

\section{The Data}

The analysis presented in the following is based on one laboratory and four different atmospheric data sets. 
In addition to accessibility reasons the latter were chosen in such a way that their environmental and meteorological characteristics differ significantly. Therefore one offshore and three onshore data sets are examined. Additionally a laboratory data set was chosen as an example of an approximately isotropic turbulent wake-flow.\\
The first data set was recorded in October 1997 near the German coastline of the North Sea in Emden at a height of $20\;m$ by means of an ultrasonic anemometer \cite{hohlen}. The wind speed was measured continuously over a period of $275$ hours. In the following this data set will be referred to as {\it On1}.\\ 
The second data set -- denoted as {\it On2} -- was obtained from a hot-wire measurement $6\;m$ above the ground (in flat terrain) \cite{kleve}. Here the wind speeds of approximately one hour were considered.\\
{\it On3} -- the third data set was recorded in a very complex terrain near Oberzeiring ($1900\; m$ above the sea level) in Austria in 2001 \cite{dewi}. The velocity was measured by means of an unltrasonic anemometer. The data consists of $255$ non-successive blocks of $4$ hours length. The choice was made in order to obtain complete and continuous data within each block.\\
The fourth data set -- referred to as {\it Off} -- was recorded during an offshore measuring campaign at Roedsand in the Danish Baltic Sea at $30\;m$ height between 1998 and 1999 \cite{rebecca}. From this period $58$ non-successive days were chosen. Again the choice was made in order to obtain complete and continuous data for each day. \\
The laboratory data -- denoted as {\it Lab} -- were obtained from a wake flow measurement in the wind tunnel of Erlangen in 1998 \cite{lueck}. A hot-wire was located $2\;m$ behind a cylinder of diameter $D=0.02\;m$ in the plane of the cylinder. Here the turbulent flow can be considered to be locally isotropic and fully developed. With a mean velocity of $\bar{u}=20.9\;ms^{-1}$ a Reynolds number of $Re=\bar{u}D\nu^{-1} \approx 30,000$ is obtained. 
Taylor and integral scale are found to be $\Theta \approx 2\cdot 10^{-4}\;s$ and $T\approx 6\cdot 10^{-3}\;s$ respectively.\\

\noindent For atmospheric data sets it is difficult to define an integral scale $T$ because of the instationarity of atmospheric velocities that causes very long-range correlations $R(\tau)\propto \; \left<u(t+\tau)u(t)\right>$ so that the integral time
\begin{eqnarray}
	T:=\int\limits_{0}^{\infty} R(\tau)\; d\tau
	\label{T}
\end{eqnarray} 
cannot be estimated properly. Only for {\it On2} an estimate of the integral time can be given because of the rather constant flow conditions during the short measuring period of about $1$ hour.\\
For {\it On2} the Taylor scale $\Theta$ can be calculated while for all other atmospheric data sets the sample frequency is too low for $\Theta$ to be resolved. 

\noindent In Table \ref{tabu} a short overview of the most important specifications of all five data sets is given. 
\begin{table}[h]
		\begin{center}
	
		\begin{tabular}{|l||c|c|c|c|c|c|}
			\hline
			$\;\;\;\;$ & ${\bf On1}$ & ${\bf On2}$ & {\bf On3} & ${\bf Off}$ & ${\bf Lab}$\\
				\hline
				\hline
				{\bf N} [1] & 3,958,874 & 20,480,000 & 14,688,000 & 25,056,000 & 12,500,000\\
				\hline
			  	{\bf blocks} [1] & 1 & 1 & 255 & 58	& 100 \\
				\hline
			 	{\boldmath$f$} [$Hz$] & 4 & 5,000 & 4 & 5 & 100,000\\
				\hline
			 	{\boldmath${\overline u}$} [$ms^{-1}$] & 3.4 & 8.3 & 6.6 & 9.6 & 20.9\\
				\hline
			 	{\boldmath$min$} [$ms^{-1}$] & 0.0 & 0.8 & 0.02 &	0.6 & 15.9\\
				\hline
			 	{\boldmath$max$} [$ms^{-1}$] & 18.1 & 18.1 & 39.0 & 36.1 & 25.5\\
				\hline
				{\boldmath$\overline{\sigma}$} [$ms^{-1}$] & 1.7 & 2.3 & 4.2 & 3.2 & 1.1\\
				\hline
			 	{\boldmath$\Theta$} [$s$] & $< 0.25$ & $\approx 0.01$ & $<0.25$ & $< 0.2$ & $\approx 2\cdot10^{-4}$\\
				\hline
			 	{\boldmath$T$} [$s$] & -- & $\approx 14$ & -- & -- & $\approx 6\cdot 10^{-3}$\\

				\hline
			
		\end{tabular}
	\end{center}
	\caption{\it The table summarizes some characteristic values of the different data sets, from top to 		bottom these are number of data points $N$, number of blocks, sample frequency $f$, mean velocity ${\overline u}$, minimum, maximum, average standard deviation ${\overline \sigma}$, Taylor scale $\Theta$ and integral scale $T$.}	
\label{tabu}
\end{table}

\section{Analysis}

\subsection{Scaling in Isotropic and Atmospheric Turbulence}	\label{sca}

\begin{figure}[h]
\caption{Scaling Exponents and Structure Functions}
\label{fig1}
\subfigure[This plot shows the scaling exponents $\zeta_{n}$ as a function of order $n$ for all five data sets. 	Additionally the linear scaling law given by Eq. (\ref{k41}) (straight line) and the non-linear laws according to Eq. (\ref{k62}) (curved line) and Eq. (\ref{sl}) (curved dashed line) are shown.] {\includegraphics[width=0.49\textwidth]{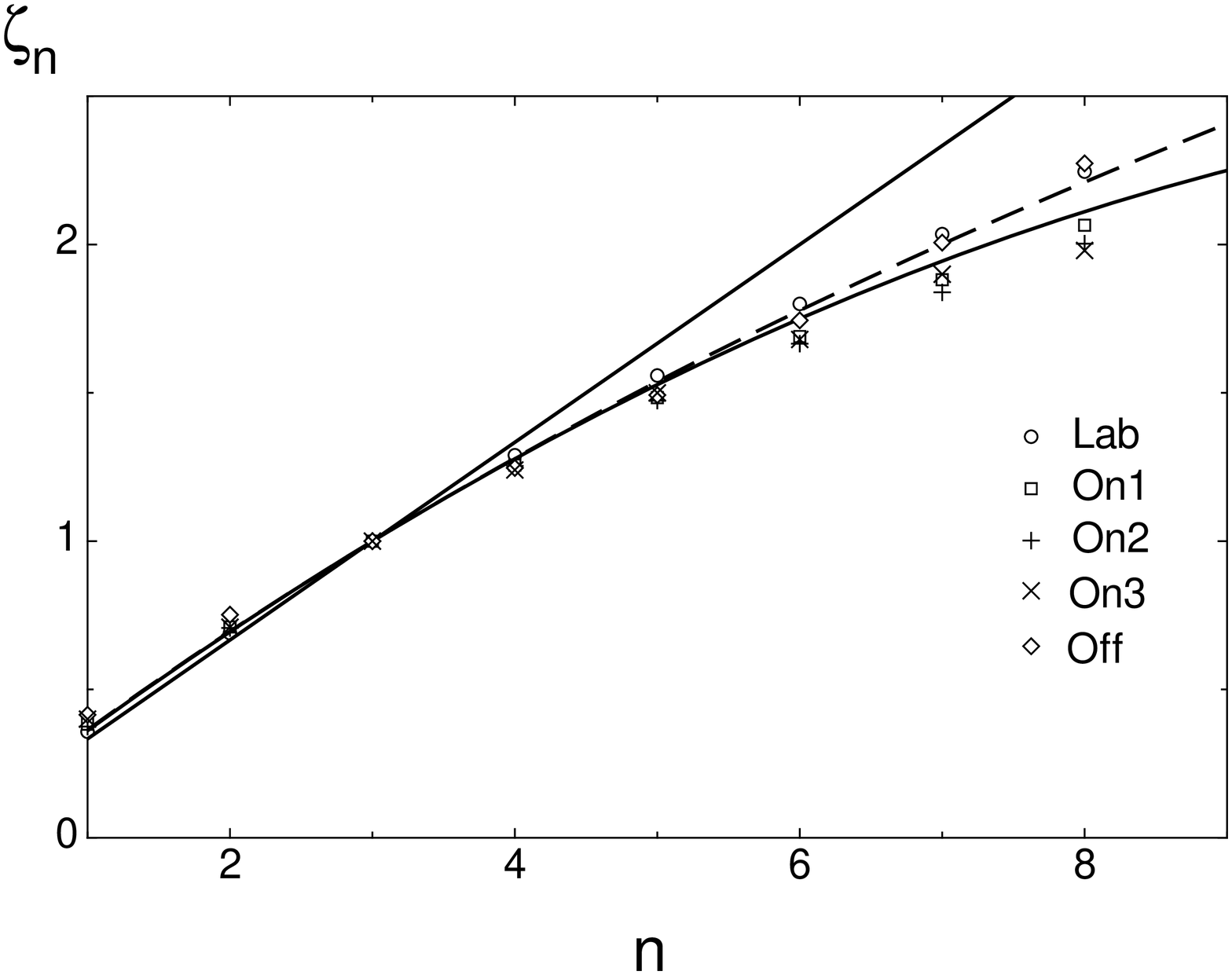}} \hfill
 \subfigure[The structure functions of order $2$, $4$ and $6$ are plotted against that of order $3$ in a double-		logarithmic presentation (vertically shifted for clarity of presentation). The open symbols belong to {\it On1}, the filled ones to {\it Lab}.] {\includegraphics[width=0.49\textwidth]{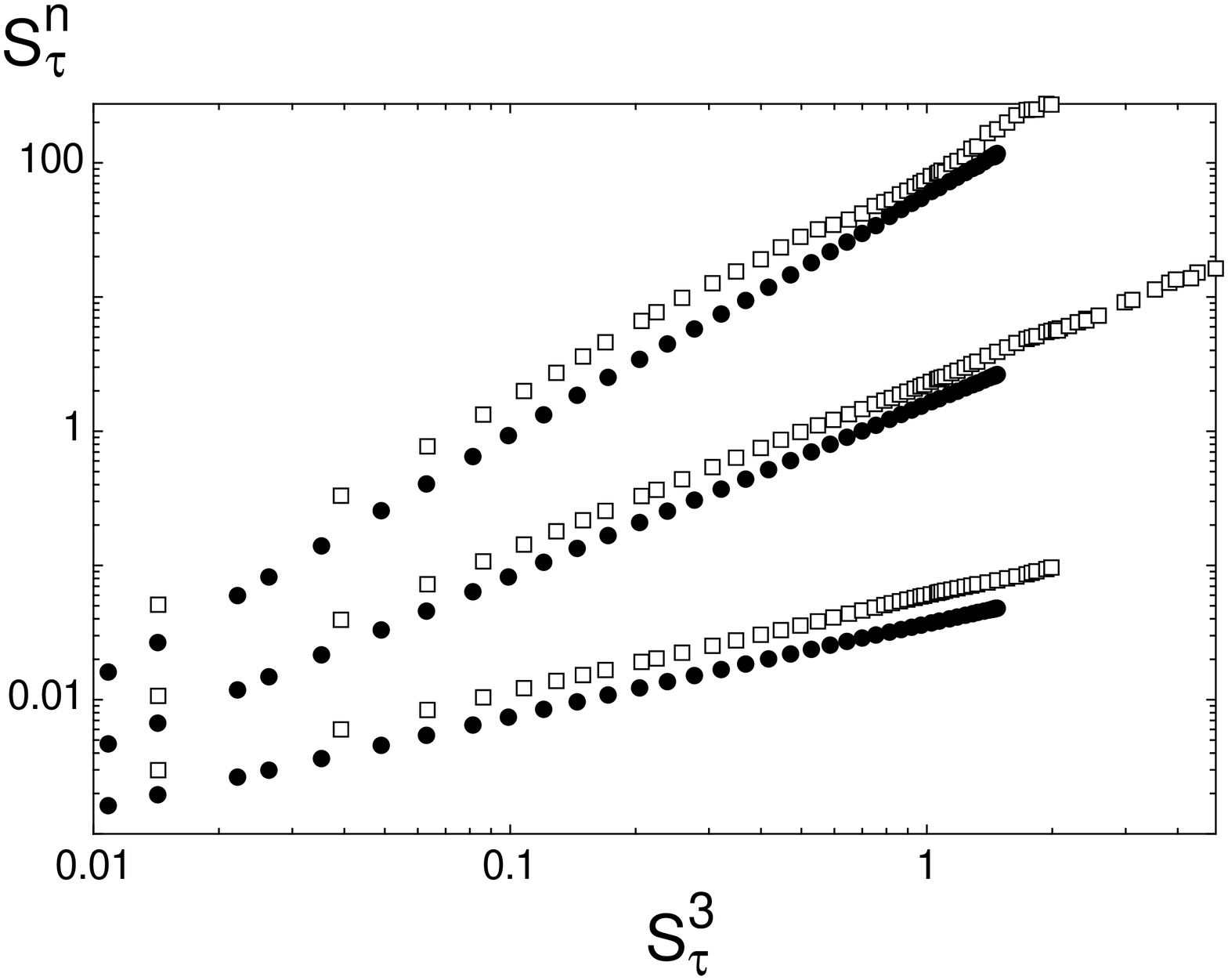}}
\end{figure}

The central assumption for turbulent velocity time series is that they have a self-similar structure (a direct consequence of scale invariance of the Navier Stokes Equation) in the {\it inertial range}. This means that within this range the disorder of velocity fluctuations has a similar structure on every scale but with a scale-dependent magnitude. To quantify this one usually calculates the (absolute) moments of increments
\begin{eqnarray}
	S_{\tau}^{n}= \left<|u_{\tau}|^{n}\right>\;\;\; ,
	\label{struc}
\end{eqnarray}
which are also called {\it structure functions} of order $n$. Instead of calculating the absolute moments one can also consider $\left<u_{\tau}^{n}\right>$ (see \cite{christoph} for a more detailed discussion). \\
In isotropic turbulence structure functions are assumed to scale as:
\begin{eqnarray}
	S_{\tau}^{n}\propto \tau^{\zeta_{n}}\;\;\; .
	\label{scale}
\end{eqnarray}
A linear (monofractal) scaling exponent $\zeta_{n}$ corresponds to a self-similar structure as proposed by Kolmogorov in 1941 \cite{k41} who found that the scaling exponent should be 
\begin{eqnarray}
	\zeta_{n}=\frac{n}{3}
	\label{k41}
\end{eqnarray}
due to dimensional reasons. Instead of this linear behaviour various experiments suggest that the scaling exponent is a non-linear function of $n$. In 1962 Kolmogorov \cite{k62} introduced the following non-linear (multifractal) exponent
\begin{eqnarray}
	\zeta_{n}=\frac{n}{3}+\frac{\mu}{18}(3n-n^{2})\;\;\; 
	\label{k62}
\end{eqnarray}
motivated by the model of a turbulent cascade with a log-normally distributed energy transfer rate. The parameter $\mu$ is called intermittency correction and is found to be close to $0.25$ \cite{mu} which corresponds to $\zeta_{6}=\frac{n}{3}-\mu=1.75$. This is in agreement with the examined data sets where $\zeta_{6}$ takes values between $1.67$ and $1.80$ as shown in Fig. \ref{fig1}a).

\noindent In \cite{she} another formula was proposed which seems to fit experimental data slightly more accurately than Eq. (\ref{k62}):
\begin{eqnarray}
	\zeta_{n}=\frac{n}{9}+2-2\left(\frac{2}{3}\right)^{n/3}\;\;\; .
	\label{sl}
\end{eqnarray}  

\begin{figure}[h]
  \begin{center}
   \includegraphics[width=8.0cm]{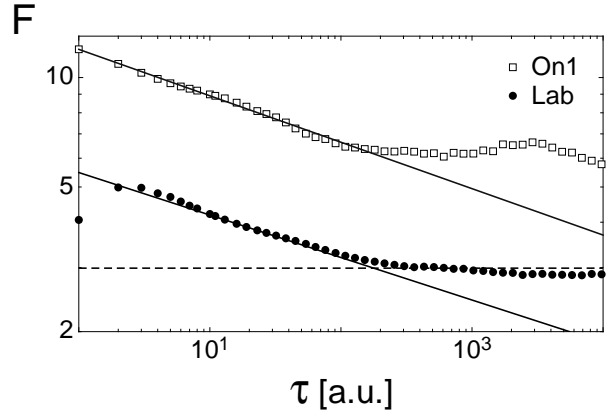}
  \end{center}
  \caption{The open symbols show the flatness $F$ of {\it On1} as a function of scale with $\tau=0.25\;s$. The filled symbols represent the flatness of {\it Lab} with $\tau=0.002\;T\approx 10^{-5}\; s$. Additionally fits according to Eq. (\ref{k62flat}) (straight lines) are drawn in with $\mu=0.29$ ({\it On1}) and $\mu=0.26$ ({\it Lab}). The horizontal dashed line marks $F=3$ (as for a Gaussian distribution).}
  \label{fig2}
\end{figure}
  
\noindent There are other models besides these two multifractal ones, e.g. \cite{lode}, \cite{yakhot}. The differences in $\zeta_{n}$ in all these models are rather small (at least for small orders) so for simplicity we will restrict following discussions to the models given by Eq. (\ref{k62}) and Eq. (\ref{sl}).\\
To estimate the dependence of $\zeta_{n}$ on $n$ one has to calculate $\zeta_{n}$ first.
The most common way to do this is to plot $log(S_{\tau}^{n})$ against $log(S_{\tau}^{3})$ -- a method referred to as  Extended Self Similarity (ESS) \cite{benzi}. The slope of the resulting line is equal to $\zeta_{n}$. This is shown in Fig. \ref{fig1}b) exemplary for data sets {\it On1} and {\it Lab}. In both cases the slopes are in quite good agreement with Eq. (\ref{k62}) and Eq. (\ref{sl}). The slopes of {\it On1} show small deviations from linear behaviour for large values. These correspond to large scales $\tau$ that might not belong to the inertial range anymore.
This already indicates that care should be taken when transferring standard analysis of isotropic to atmospheric turbulence.\\ 
The difference between isotropic and atmospheric turbulence statistics becomes more obvious when calculating the flatness $F$. Assuming inertial range scaling according to Eq. (\ref{struc}) and (\ref{scale}) the flatness should scale as well and is given by:
\begin{eqnarray}
	F\; :=\; \frac{\left<u_{\tau}^{4}\right>}{\left<u_{\tau}^{2}\right>^{2}} \; \propto \; \tau^{\zeta_{4}-2\zeta_{2}} \;\;\; .
	\label{flat}
\end{eqnarray}
If Eq. (\ref{k62}) is a suitable description\footnote{Eq. (\ref{sl}) and other multifractal models yield very similar results because such low-order exponents as $\zeta_{4}$ and $\zeta_{2}$ are quite indistinguishable from each other.} the flatness scales according to:
\begin{eqnarray}
	F \; \propto \; \tau^{-4 \mu/9} \;\;\; ,
	\label{k62flat}
\end{eqnarray}
as is easily shown inserting Eq. (\ref{k62}) into Eq. (\ref{flat}). As shown in Fig. \ref{fig2} the measured flatness of the data shows this scaling behaviour but the absolute values of $F$ are very different for different data sets. While for the {\it Lab} data set flatness approaches $F\approx3$ for large $\tau$ it saturates at $F\approx6$ for the {\it On1} data set. The flatness of {\it On3} and {\it Off} saturates at $F\approx 5$ while for {\it On2} it goes down to $3.5$.\\

\noindent Calculating the flatness of a variable $x$ is often done to estimate the shape of the PDF $p(x)$. A Gaussian distribution has flatness $3$. Deviations from this value can be taken as a hint for a non-Gaussian shape of PDFs. In the following the increment PDFs of the given data sets will be examined.

\subsection{Probability Density Functions in Turbulence}	\label{prob}

Alternatively to the analysis of scaling exponents one can directly investigate the PDFs of velocity increments $p(u_{\tau})$. The scaling behaviour as well as all moments (including derived quantities such as flatness or skewness) are immediately given if the distributions on every scale are known.\\
In principle the knowledge of all moments $S_{\tau}^{n}$ and the knowledge of the PDFs should contain the same information as can be seen by means of the characteristic function $\varphi(k)$ -- which is the Fourier transform of the PDF -- defined as:
\begin{eqnarray}
	\varphi(k) := \sum_{n=1}^{\infty} S_{\tau}^{n} \frac{(ik)^{n}}{n!}  \;\;\; .
	\label{phi}
\end{eqnarray}
Nevertheless the relation between moments and PDFs is not unique. In \cite{sornette} it is pointed out that two different PDFs can have exactly the same moments.\\
From many experiments of isotropic turbulence it is well known that the shape of a PDF changes with scale. Going from larger to smaller scales the distributions become more and more heavy-tailed while for $\tau \geq T$ a Gaussian distribution is obtained. This scale-dependent shape corresponds to non-linear scaling exponents -- as introduced in Eq. (\ref{k62}) and Eq. (\ref{sl}) -- while a linear behaviour $S_{\tau}^{n}=a_{n}\tau^{\alpha n}=a_{n}\beta^{n}$ according to Eq. (\ref{k41}) leads to a constant shape. This can be seen by means of the characteristic function in Eq. (\ref{phi}) that stays the same for a linear exponent only $k$ is rescaled according to $\tilde{k}=\beta k$.\\
The analysis of scaling exponents thus focusses on the change in shape of distributions while the shape itself is of minor interest and could even be determined wrongly as shown at the end of the last chapter.
Therefore we will henceforth focus on the analysis of PDFs.

\begin{figure}[h]
	\caption{Probability Density Distributions}
	\label{fig3}
	\subfigure[\label{fig3a}Symbols represent normalized PDFs (with their scale-dependent standard deviation 				$\sigma_{\tau}=\sqrt{\left<u_{\tau}^{2}\right>}$) of the {\it Lab} data set. From top to bottom $\tau$ takes the values: $0.01\;T$, $0.05\;	T$, $0.2\;T$ and $1.0\;T$ where $T\approx 0.005\;s$ denotes the integral time. Straight lines correspond to a fit of the 		distributions according to Eq. (\ref{castaing}).]	{\includegraphics[width=0.48\textwidth]{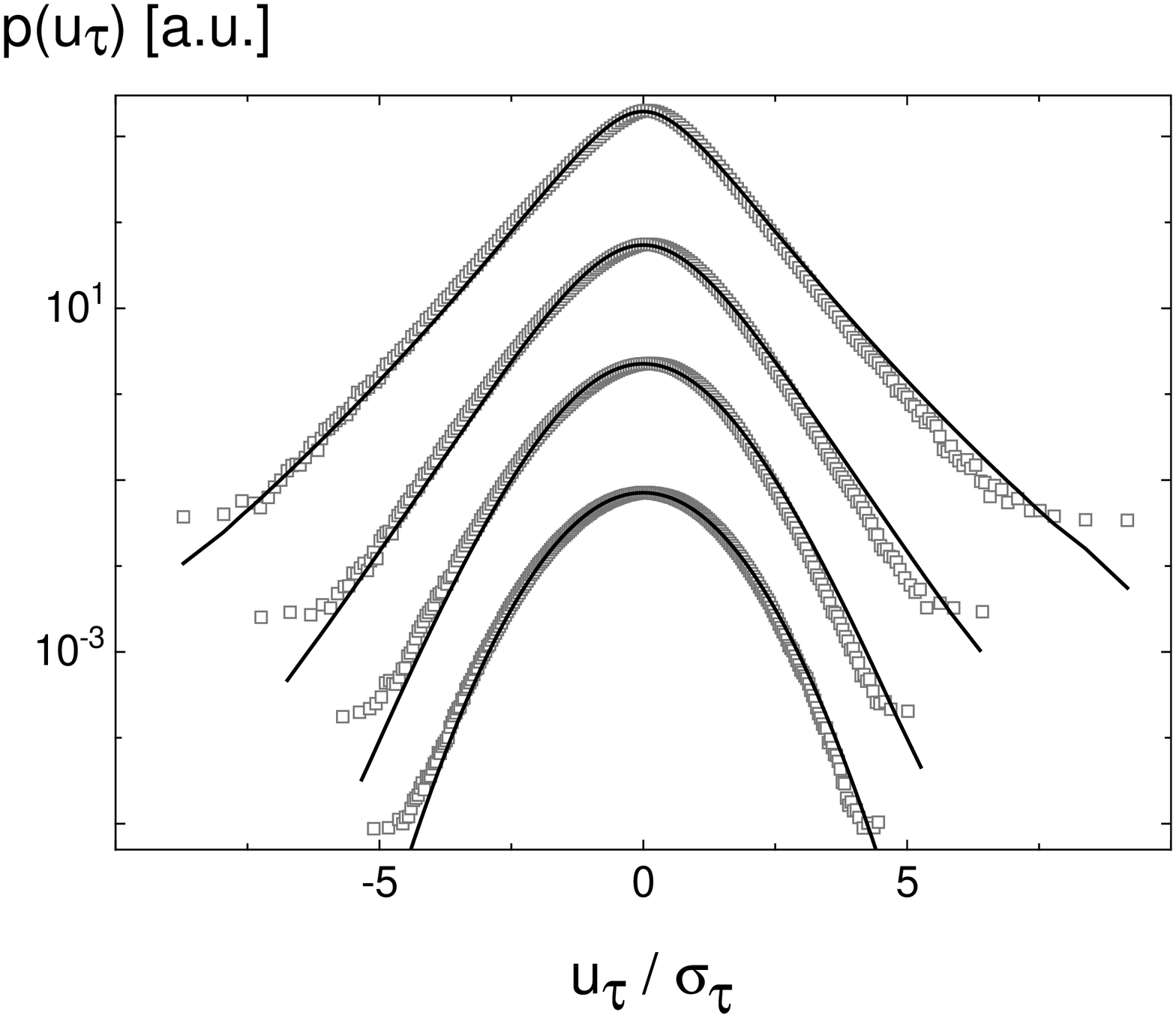}}	
	 \subfigure[\label{fig3b} Symbols represent normalized PDFs of the atmospheric data set {\it On1} with $\tau=0.5\;s$, $2.5\;s$, 	$25\;s$, $250\;s$ and $4000\;s$. Straight lines correspond to a fit of the distributions according to Eq. (\ref{castaing}). All 	graphs -- in (a) and (b) -- are plotted in a semi-logarithmic presentation and are shifted against each other in vertical 		direction for clarity of presentation.] 	{\includegraphics[width=0.48\textwidth]{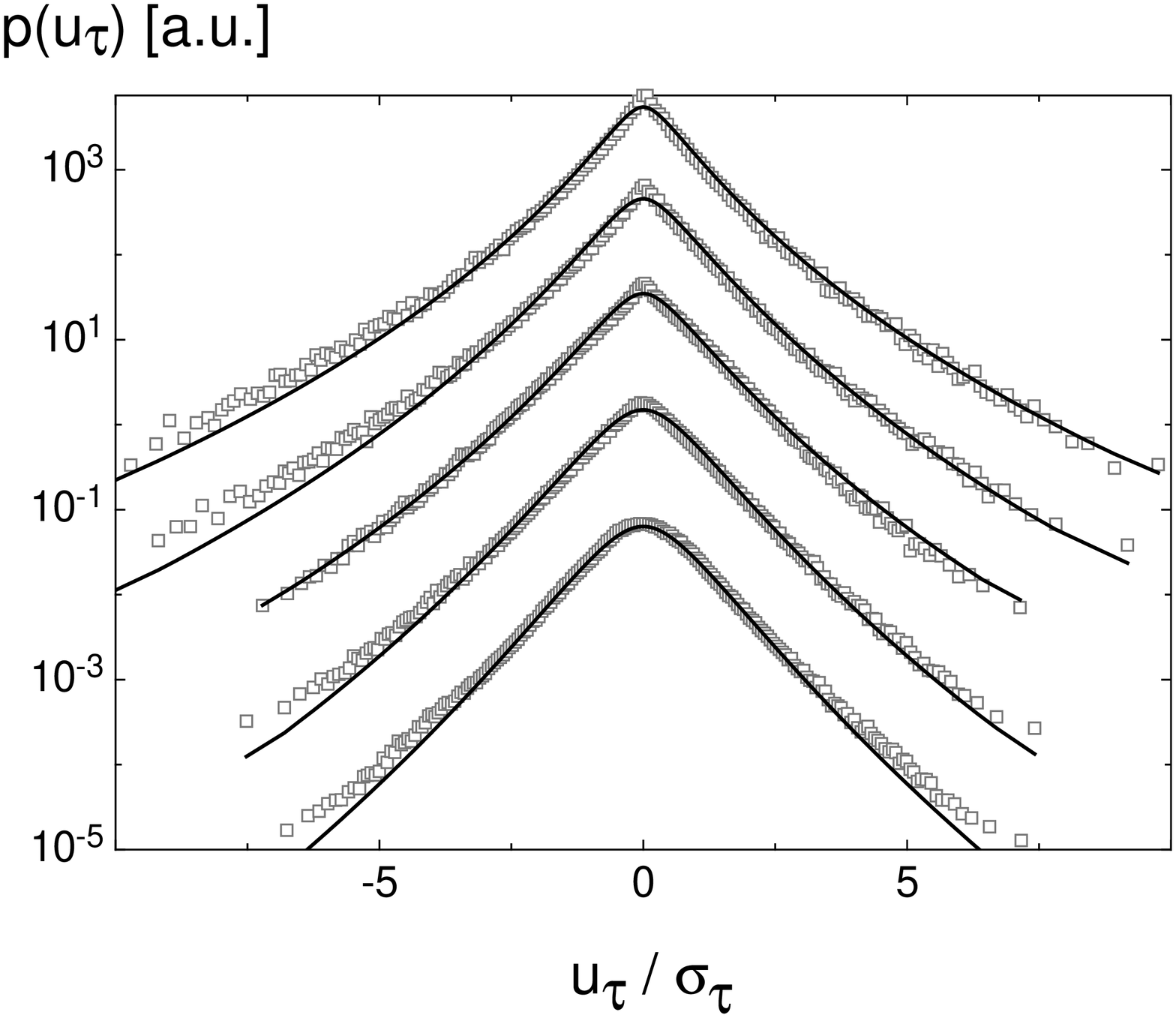}}
\end{figure}
  
\noindent In accordance with Eq. (\ref{k62}) (that was derived from the assumption of a log-normally distributed energy transfer rate) B. Castaing et al. \cite{castaing} introduced a model in which the increment distribution $p(u_{\tau})$ is interpreted as a superposition of Gaussian ones $p(u_{\tau}|\sigma)$ with standard deviation $\sigma$. The standard deviation itself is distributed according to a log-normal distribution $f(\sigma)$. The increment distribution thus reads 
\begin{eqnarray}
	p(u_{\tau}) &=& \int\limits_{0}^{\infty} d\sigma \;\; p(u_{\tau}|\sigma) \cdot f(\sigma) \nonumber \\
	&=& \int\limits_{0}^{\infty} d\sigma \; \frac{1}{\sigma\sqrt{2\pi}} exp\left[-\frac{u_{\tau}^{2}}{2\sigma^{2}}\right] 	\cdot \frac{1}{\sigma \lambda \sqrt{2\pi}} exp\left[-\frac{ln^{2}(\sigma/\sigma_{0})}{2\lambda^{2}}\right] \;\;\;
	\label{castaing}
\end{eqnarray} 
and will henceforth be referred to as the {\it Castaing distribution}.\\
Two parameters enter this formula, namely $\sigma_{0}$ and $\lambda^{2}$. The first is the median of the log-normal distribution, the second its variance. The latter determines the form (shape) of the resulting distribution $p(u_{\tau})$ and is therefore called {\it form parameter}. On one side the larger $\lambda^{2}$ becomes the broader the log-normal distribution and the broader the range of $\sigma$ that contributes to the integral in Eq. (\ref{castaing}). On the other side the range of $\sigma$ becomes smaller and smaller with decreasing form parameter. In the limit of vanishing $\lambda$ the log-normal becomes a delta distribution
\begin{eqnarray}
	\lim\limits_{\lambda \rightarrow 0} \;\; \left( \frac{1}{\sigma \lambda \sqrt{2\pi}} exp\left[-\frac{ln^{2}(\sigma/\sigma_{0})}{2\lambda^{2}}\right]\right) \; = \; \delta(\sigma-\sigma_{0}) \;\;\; ,
	\label{delta}
\end{eqnarray}
so that $p(u_{\tau})$ is reduced to a Gaussian distribution with variance $\sigma_{0}^{2}$.\\
With a proper choice of the form parameter the PDFs $p(u_{\tau})$ can well be fitted as it is shown in Fig. \ref{fig3}a) and Fig. \ref{fig3}b). In Fig. \ref{fig3}a) -- where the {\it Lab} data set is presented -- the expected change of shape from intermittent to Gaussian distributions with increasing scale is clearly seen.\\

\begin{figure}[h]
  \begin{center}
   \includegraphics[width=7.0cm]{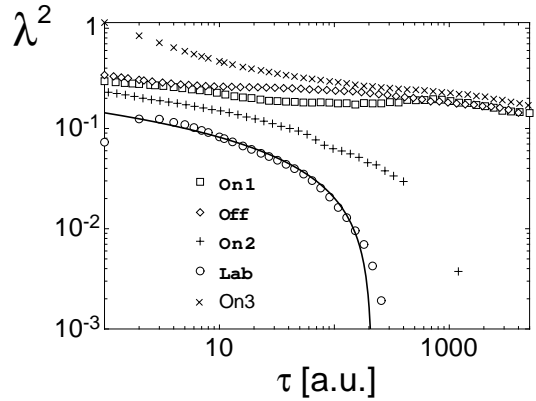}
  \end{center}
  \caption{  \label{fig4}The form parameter for the data sets {\it On3}, {\it On1}, {\it Off}, {\it On2} and {\it Lab} are shown in double-logarithmic presentation. In that order $\tau=1$ corresponds to $0.25\;s$, $1\;s$, $1\;s$, $0.02\;s$ and $10^{-5}s$ ($0.002\;T$). The straight line represents a fit according to Eq. (\ref{lama}).}

\end{figure}

\noindent In contrast to this behaviour the PDFs of {\it On1} in Fig. \ref{fig3}b) look totally different. They are much more intermittent and do not approach a Gaussian distribution even for very large scales. For scales larger than about $25\;s$ the shape remains rather constant. This is in accordance with the finding of the slow decrease of flatness shown in Fig. \ref{fig2} because in the log-normal model flatness is linked to the form parameter \cite{beck} according to 
\begin{eqnarray}
	\lambda^{2}\propto ln\left(\frac{F}{3}\right)\;\;\; . 
	\label{flatlam}
\end{eqnarray}
In this sense constant flatness larger than $3$ corresponds to constant $\lambda^{2}$ larger than $0$ and thus to scale-independent intermittent distributions as shown in Fig. \ref{fig2} and Fig. \ref{fig3}b) exemplary for the {\it On1} data set. The $\lambda^{2}$-values for all data sets are illustrated in Fig. \ref{fig4}.\\
For isotropic turbulence scale-independent distributions occur for large scales as well but are always found to be Gaussian \cite{gauss} for scales larger than the integral time $T$. From a mathematical point of view these {\it stable} Gaussian PDFs are the result of a stable stochastic Gaussian process. The  other class of stable distributions are the so-called L\'evy distributions -- characterised by power-law tails -- which are the result of a fractional stochastic process. The observed atmospheric increment PDFs show robust (stretched) exponential tails that decay faster than a power-law and slower than a Gaussian distribution. So the question arises whether the atmospheric PDFs can be explained as a fractional process or as a superposition of different Gaussian processes in analogy to the {\it Castaing distribution}.
To decide this the concept of stability and the connection to increment analysis should briefly be introduced.\\

\subsubsection{Stable Distributions}

Consider the sum $s_{m}=\sum\limits_{i}^{m} x_{i}$ of $m$ independent and identical distributed (i.i.d.) variables. The variables should be distributed according to a PDF $p(x)$ and the PDF of the sum-variable is $\hat{p}(x)$. The distribution $p$ (or $\hat{p}$ respectively) is then called {\it stable} if for large $m$ ($m\rightarrow \infty$)
\begin{eqnarray}
	\hat{p}(x')\; dx'\;=\; p(x)\;dx\;\;\;\; with \;\;\;\; x'=Ax+B
	\label{stab}
\end{eqnarray}
is fulfilled \cite{sornette}. This means that for sufficiently large $m$ the shape of the distribution does not change as $m$ increases.\\
Transferred to increment analysis an increment over a scale $\tau$ can be identified as variable $x$ and the increment of a larger scale $m\tau$ as the sum-variable $s_{m}$. It is immediately shown that 
\begin{eqnarray}
	u_{m\tau}(t)=\sum\limits_{i=0}^{m-1} u_{\tau}(t+i\tau)\;\;\; ,
	\label{ninc}
\end{eqnarray}
which means that a large increment can be expressed as the sum of smaller ones. When the PDFs of large and small increments are the same this indicates a stable PDF.

The most famous stable distribution is the Gaussian one. Beside this P. L\'evy \cite{levy} showed that there exists a whole class of stable distributions. Restricting to  symmetric distributions their characteristic functions read: 
\begin{eqnarray}
	\varphi(k) = exp\left[i\gamma k - c|k|^{\alpha}\right] \;\;\;\;\;\; ; 0< \alpha \leq 2.
	\label{levy1}
\end{eqnarray}
For asymmetric distributions the characteristic function becomes more complicated (e.g. \cite{paul}). The analytical form of the corresponding PDFs is only known for some special cases (e.g. for $\alpha=1$ it is the Cauchy distribution) but their asymptotic behavior is always known and given by:
\begin{eqnarray}
	p(x)\; \propto \; C\;|x|^{-(1+\alpha)} \;\;\;;\; x\gg1\;\;\;.
	\label{levy2}
\end{eqnarray}
This algebraic decay of tails means that all higher order moments larger than order $\alpha$ do not exist. Generally, distributions with tails decaying faster than $\propto |x|^{-3}$ (defined variance) can only converge to a Gaussian while slower decaying tails indicate that they can only converge to a L\'evy stable law. 

As already mentioned the examined atmospheric PDFs show a faster than algebraic decay (compare Fig. \ref{fig3} b) and Fig. \ref{fig8} a), b), c), d)) so it is expected that they converge to Gaussian statistics for large scales. Therefore the observed robust intermittency should be explained by mixing different Gaussian distributions rather than by a fractional stochastic process.  
\clearpage

\subsection{Superposition Model for Atmospheric Turbulence}		\label{super}

\begin{figure}[h]
  \begin{center}
   \includegraphics[width=7cm]{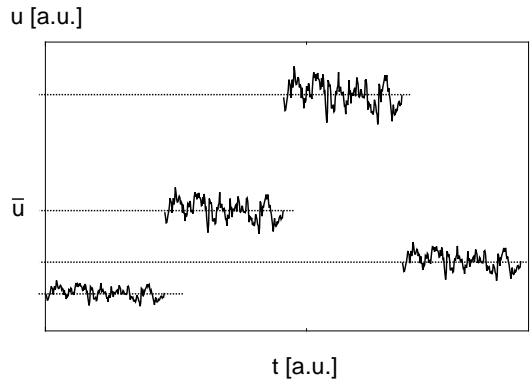}
  \end{center}
   \caption{ \label{fig5}Illustration of different mean velocity intervals. Within these intervals statistics should be the same as for isotropic turbulence. The magnitude of variations (standard deviation) grows with mean velocity according to Eq. (\ref{lin}).}
\end{figure}

In \cite{us} it was proposed to take the instationarity of long atmospheric time series into account. In this sense the observed intermittent form of PDFs for all examined scales are found to be the result of mixing statistics belonging to different flow situations. These are characterized by different mean velocities as schematically illustrated in Fig. \ref{fig5}. 
When the analysis by means of increment statistics is conditioned on periods with constant mean velocities results are found to be very similar to those of isotropic turbulence. This can be set in analogy to the {\it Castaing distribution} that interprets intermittent PDFs as a superposition of intervals with different standard deviations. \\

\noindent Thus we propose a model that describes the robust intermittent atmospheric PDFs as a superposition of those of isotropic turbulent subsets that are denoted with $p(u_{\tau}|\bar{u})$ and given by Eq. (\ref{castaing}). Knowing the distribution of the mean velocity $h(\bar{u})$ the PDFs become:
\begin{eqnarray}
	p(u_{\tau}) 	=  \int\limits_{0}^{\infty} d\bar{u} \;\;	h(\bar{u})\cdot p(u_{\tau}|\bar{u}) \;\;\; .
	\label{sumsup}
\end{eqnarray}

\begin{figure}[h]
  \begin{center}
   \includegraphics[width=8cm]{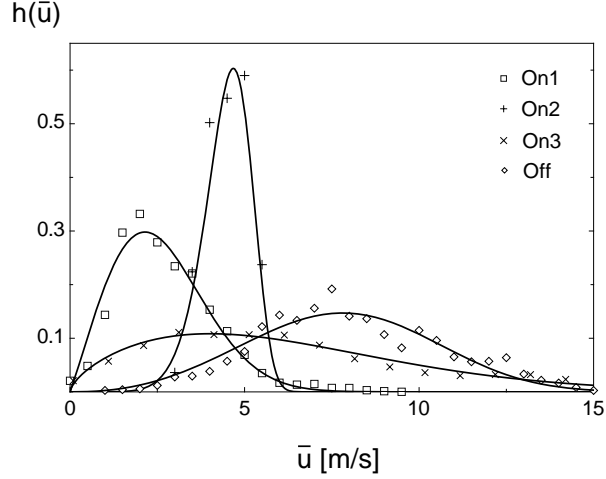}
  \end{center}
   \caption{Symbols represent measured mean velocity distributions (averaged over $10\;min$) of the four atmospheric data sets {\it On1}, {\it On2}, {\it On3} and {\it Off}. Solid lines are fits according to Eq. (\ref{weibull}). The parameters are: $k=2.0,\;7.8,\;1.8,\;3.3$ and $A=2.9,\;4.8,\;7.1,\;8.7$ for {\it On1}, {\it On2}, {\it On3} and {\it Off} respectively.}
  \label{fig6}
\end{figure}
  
\begin{figure}[h]
	\caption{Standard Deviation as a Function of $\bar{u}$ and $\tau$}	
	\label{fig7}
	\subfigure[Standard deviation $\sigma_{0}$ as a function of $\bar{u}$ and $\tau$ -- exemplary for 	the {\it On1} 	data set.]
 	{\label{fig7a}\includegraphics[width=0.49\textwidth]{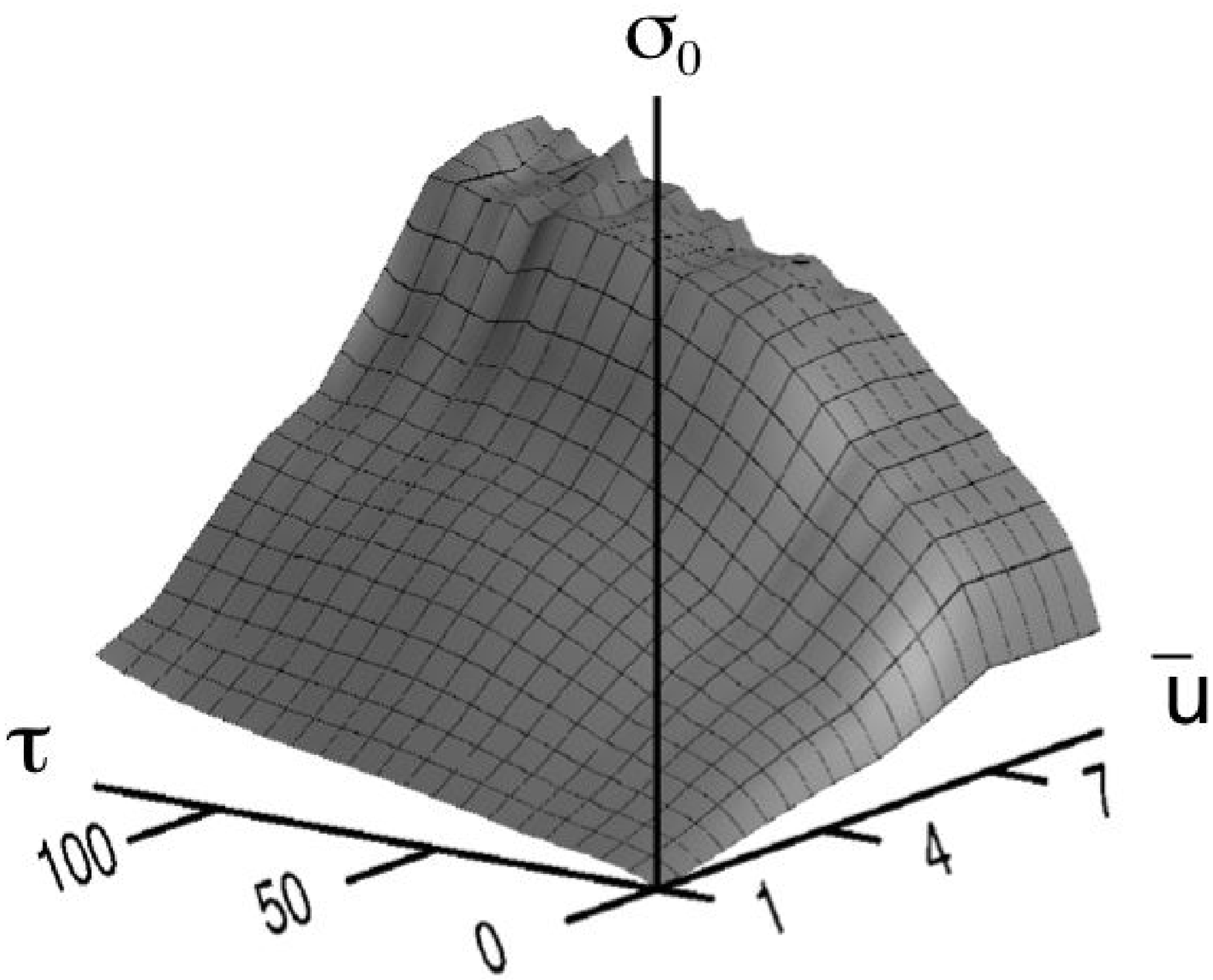}}\hfill		
 	\subfigure[Standard deviation $\sigma_{0}$ only as a function of $\bar{u}$ (at fixed $\tau$) -- 		exemplary 	for 		$\tau=1\;s$ and $\tau=50\;s$.] 
 	{\label{fig7b}\includegraphics[width=0.41\textwidth]{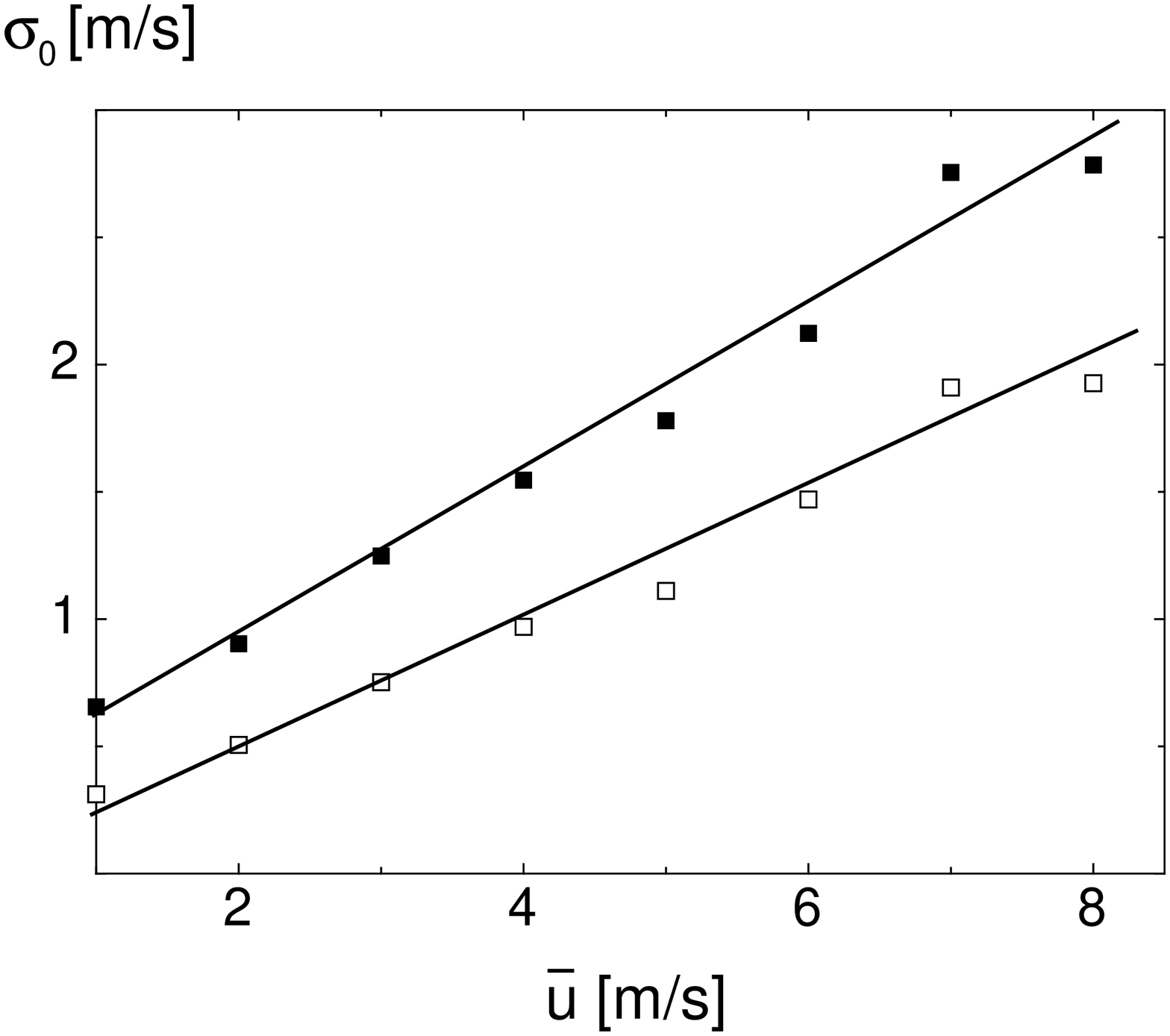}}
\end{figure}

\noindent To find a suitable averaging time defining the mean value $\bar{u}$ is a non-trivial problem due to the lack of a distinct mesoscale gap and is not the concern of the present paper (see e.g. \cite{trevino}).\\
As a first approximation we fix the averaging time to be $10$ minutes. In \cite{us} it is shown that with this choice a partition into isotropic sequences is achieved. Furthermore we assume $h(\bar{u})$ to be a Weibull distribution: 
\begin{eqnarray}
	h(\bar{u}) =  \frac{k}{A} \left(\frac{\bar{u}}{A}\right)^{k-1}  exp\left[-\left(\frac{\bar{u}}{A}\right)^{k}\right]\;\;\; .
	\label{weibull}
\end{eqnarray}	
Both assumptions are well established in meteorology \cite{burton}. In Fig. \ref{fig6} it is shown that a Weibull distribution is a good representation of $h(\bar{u})$.\\ 
Inserting Eq. (\ref{weibull}) and Eq. (\ref{castaing}) into Eq. (\ref{sumsup}) the following expression for atmospheric PDFs is obtained:
\begin{eqnarray}
	p(u_{\tau}) &=& \frac{k}{2\pi A^{k}}\int\limits_{0}^{\infty}d\bar{u}\int\limits_{0}^{\infty} d\sigma\;\;\bar{u}^{k-1}\; exp\left[ -\left(\frac{\bar{u}}{A}\right)^{k}\right] \nonumber \\ 	&\times&			\frac{1}{\lambda\sigma^{2}}\;exp\left[ -\frac{u_{\tau}^{2}}{2\sigma^{2}}\right] exp\left[ -\frac{ln^{2}(\sigma/\sigma_{0})}{2\lambda^{2}}\right] \;\;\; .
	\label{castaing2}
\end{eqnarray}
Parameters $A$ and $k$ play a similar role as $\sigma_{0}$ and $\lambda^{2}$ in the {\it Castaing distribution}. \\

\noindent Next we briefly discuss the meaning of the two parameters $A$ and $k$. 
The former is closely related to the expectation value ($\left<\bar{u}\right>=A\cdot \Gamma(1+k^{-1})$) of the mean velocity. Parameter $k$ determines the form (shape) of the distribution. Small values of $k$ correspond to a broad distribution where many different mean velocities contribute significantly to the integral in Eq. (\ref{castaing2}). Large $k$-values however prevent this contribution and the distribution is closely centered around $A$. In the limit of very large $k$ a delta distribution is obtained:
\begin{eqnarray}
	 \lim\limits_{k\rightarrow \infty}\;\; \left( \frac{k}{A^{k}}\; \bar{u}^{k-1} \; exp\left[-\left(\frac{\bar{u}}{A}\right)^{k}\right]\right)  \;=\; \delta(\bar{u}-A) \;\;\; .
	\label{delta2}
\end{eqnarray}
So for very large $k$-values Eq. (\ref{castaing2}) converges to Eq. (\ref{castaing}) i.e. to isotropic turbulence. For additionally large $\tau$ and accordingly vanishing $\lambda^{2}$ one finally gets
\begin{eqnarray}
		\lim\limits_{k \rightarrow \infty} \; \lim\limits_{\tau \rightarrow \infty} \; p(u_{\tau}) = \frac{1}{\sigma_{0}\sqrt{2\pi}} exp\left[-\frac{u_{\tau}^{2}}{2\sigma_{0}^{2}} \right] 
	\label{2lim}
\end{eqnarray}
which is a pure Gaussian density with variance $\sigma_{0}^{2}(A)$. For small $k$-values the distributions remain intermittent even for large scales. Here intermittency is caused by the mixing of periods of different mean velocities.\\
    
\noindent To apply Eq. (\ref{castaing2}) to experimental data one has to know the parameters $A$, $k$, $\sigma_{0}$ and $\lambda^{2}$. Parameters $A$ and $k$ can directly be estimated from data by fitting a Weibull distribution given in Eq. (\ref{weibull}) to the measured distribution of $\bar{u}$. This is illustrated in Fig. \ref{fig6} for all examined atmospheric data sets.\\ 
Next $\lambda^{2}=\lambda^{2}(\tau, \bar{u})$ and $\sigma_{0}=s_{0}(\tau, \bar{u})$ have to be known. From isotropic turbulence it is well-established that the form parameter decreases monotonously in scale. Scaling of flatness according to Eq. (\ref{flat}) together with Eq. (\ref{flatlam}) directly leads to a logarithmic depenency of $\lambda^{2}$ on $\tau$:
\begin{eqnarray}
	\lambda^{2} = a_{\bar{u}} - b_{\bar{u}} \cdot ln(\tau) \;\;\; .
	\label{lama}
\end{eqnarray}
To confirm this in Fig. \ref{fig4} the form parameter of the {\it Lab} data set is fitted with Eq. (\ref{lama}). For a deeper discussion on the scale dependence of $\lambda^{2}$ see \cite{chabaudermann}.\\
The dependence of $\lambda^{2}$ on $\bar{u}$ is small (e.g. \cite{castaing} found a slower decay of $\lambda^{2}$ for increasing Reynolds numbers $Re\propto \bar{u}$). As a first approximation we will disregard it.\\
The parameter $\sigma_{0}$ -- the most probable standard deviation -- generally grows with scale and mean velocity. The scale dependence drops out automatically when considering normalized PDFs as it is done here. For the $\bar{u}$-dependence we assume a linear relation
\begin{eqnarray}
	s_{0}=b_{\tau}\cdot \bar{u} \;\;\; ,
	\label{lin}
\end{eqnarray}
which is supported from the measured standard deviation $\sigma_{0}$, shown in Fig. \ref{fig7}.\\
Thus, to apply Eq. (\ref{castaing2}) we need to calculate $A$ and $k$ from measured mean velocity distributions. Additionally we consider a logarithmic dependence of $\lambda^{2}$ on $\tau$ and a linear one for $\sigma_{0}$ on $\bar{u}$. Given that we are interested in a preferably simple reconstruction of PDFs by means of Eq. (\ref{castaing2}) we will disregard the more complex parameter relations here.\\

\begin{figure}[h]
	\caption{Atmospheric Probability Density Distributions: Symbols represent the normalized PDFs of the four atmospheric data sets in semi-logarithmic presentation. Straight lines correspond to a fit of distributions according to Eq. (\ref{castaing2}). All graphs are vertically shifted against each other for clarity of presentation.}
	\label{fig8}
	\subfigure[{\it Off}: From top to bottom $\tau$ takes the values: $0.2\;	s$, $10\;s$, $20\;s$, $200\;s$ and $2000\;s$. ]	{\includegraphics[width=0.49\textwidth]{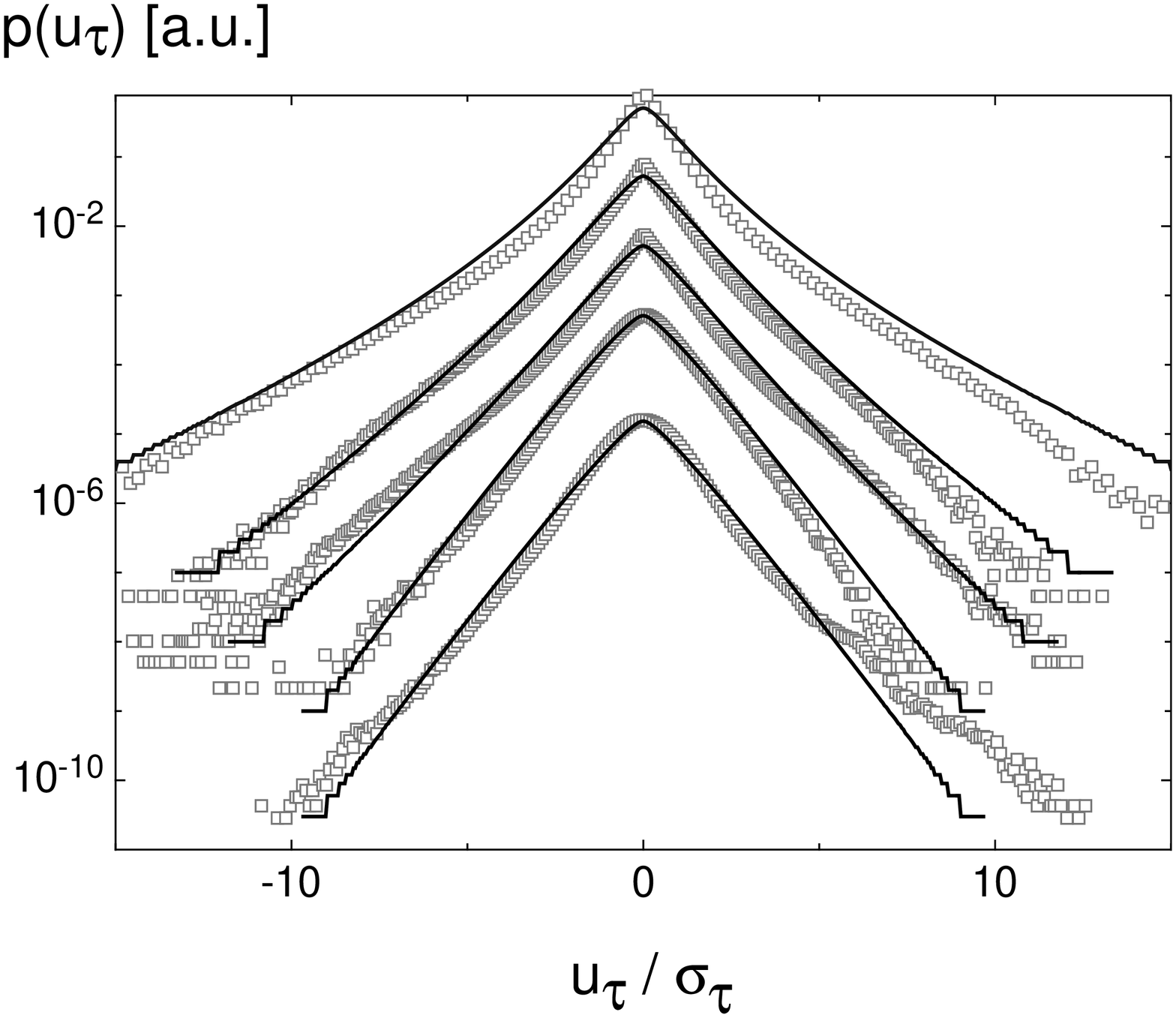}}
	 \subfigure[{\it On1}: From top to bottom $\tau$ takes the values: $0.5\;s$, $2.5\;s$, $25\;s$, 	$250\;s$ and $4000\;s$.] 
 	{\includegraphics[width=0.49\textwidth]{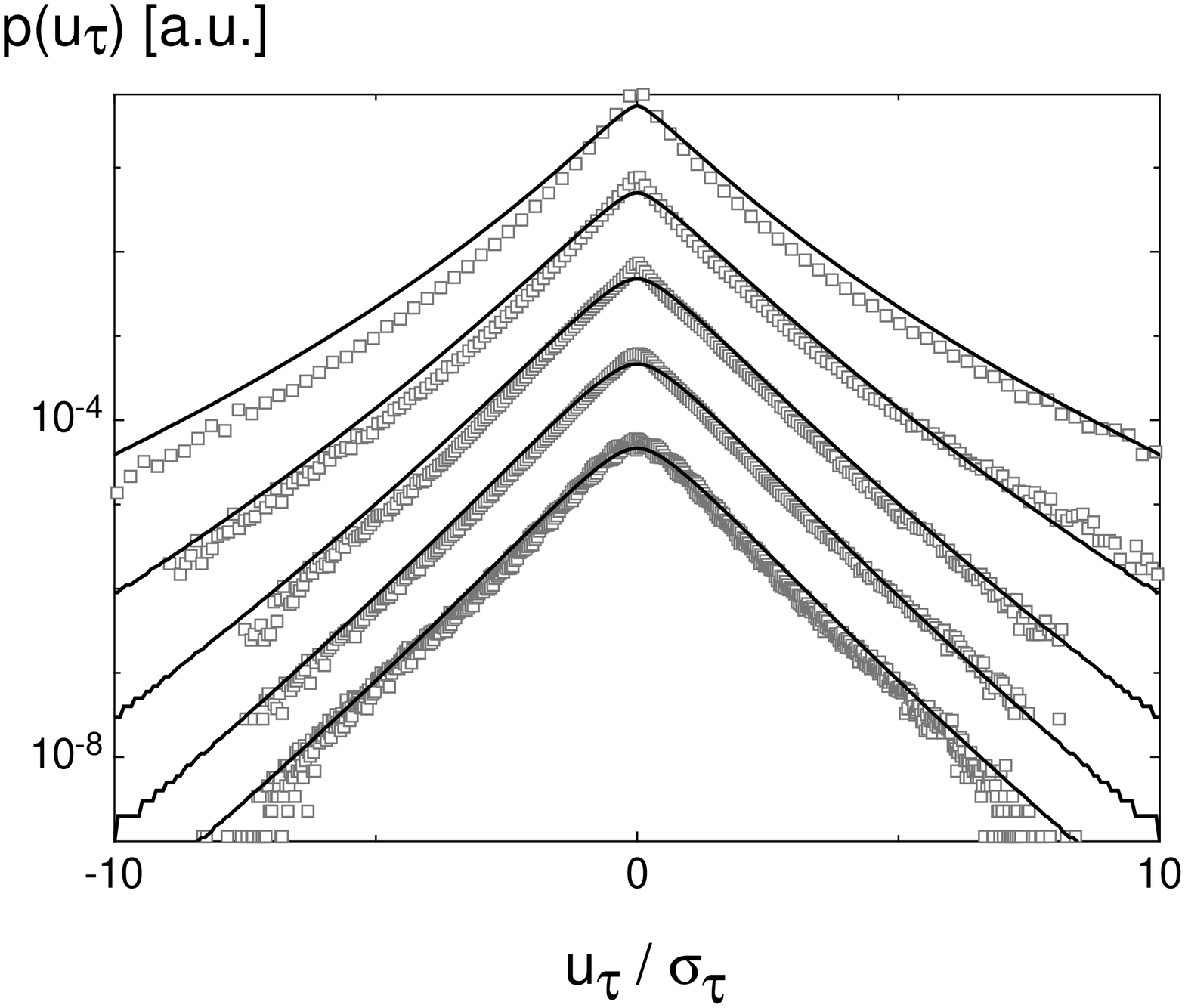}}
	\subfigure[{\it On2}: From top to bottom $\tau$ takes the values: $2\;		ms$, $20\;ms$, $200\;ms$ and $2000\;ms$.]	{\includegraphics[width=0.49\textwidth]{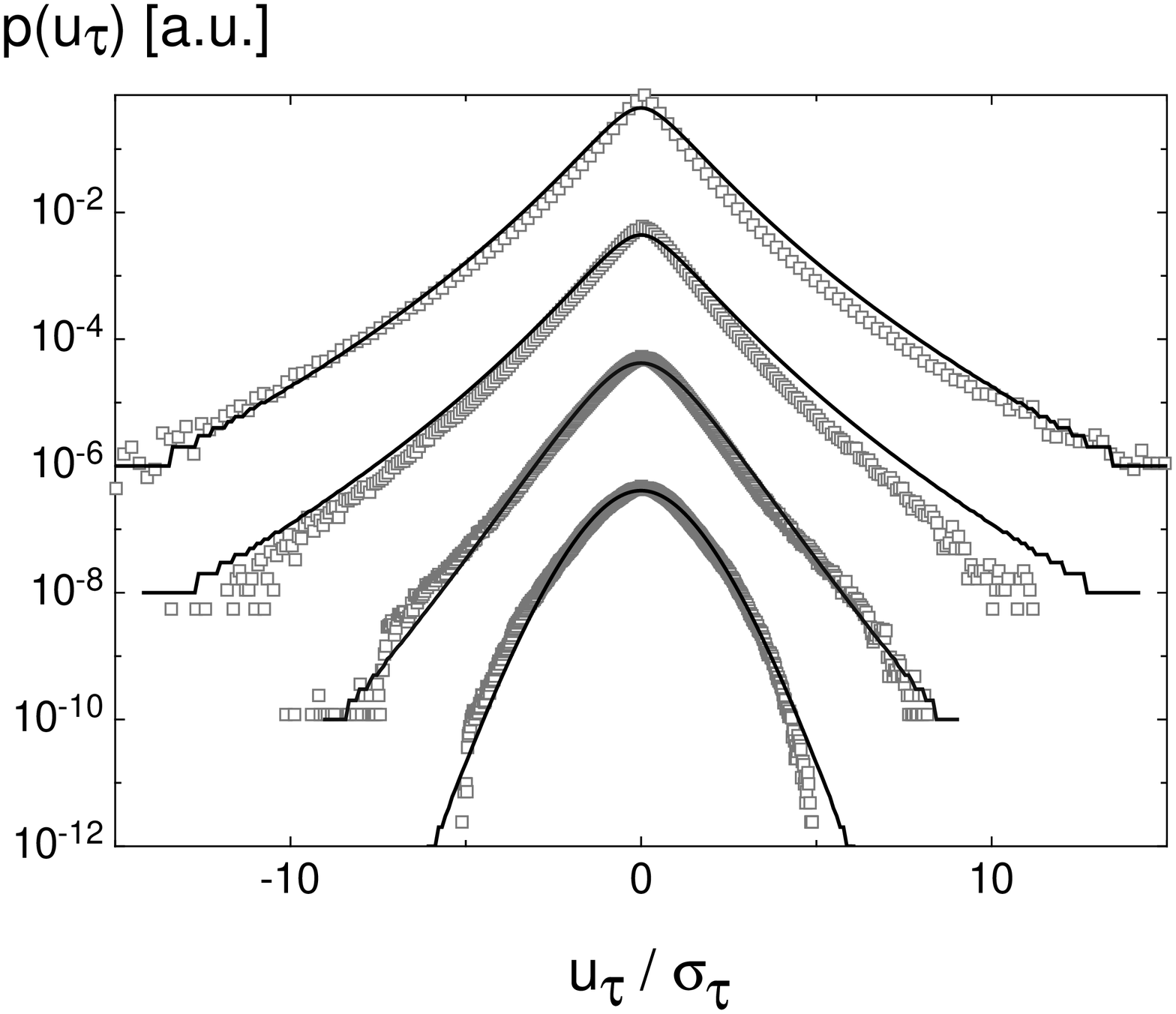}}
	 \subfigure[{\it On3}: From top to bottom $\tau$ takes the values: $0.25\;	s$, $2.5\;s$, $25\;s$ and $250\;s$.] 
 	{\includegraphics[width=0.49\textwidth]{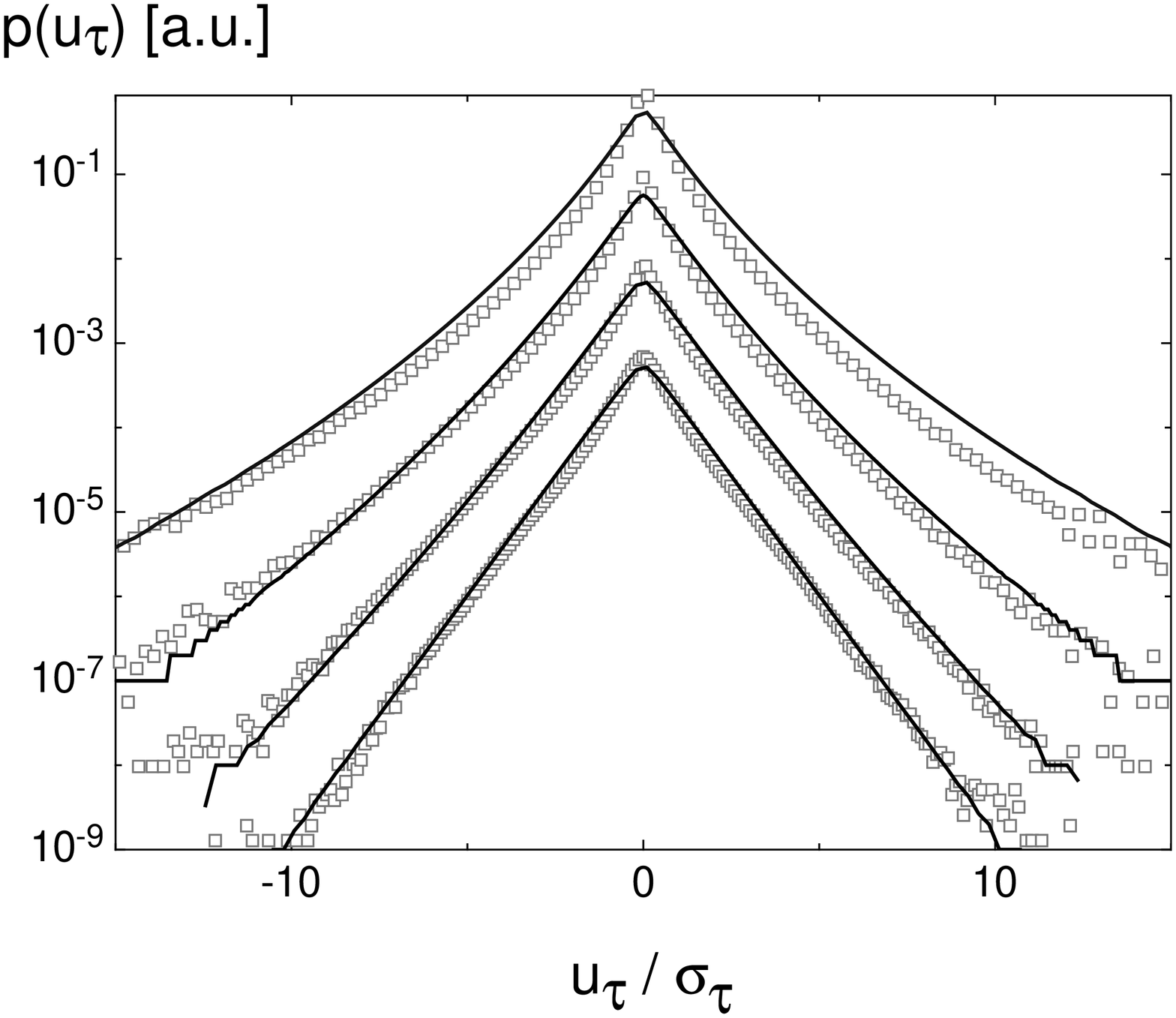}}
\end{figure}

\noindent With these parameters atmospheric increment PDFs can be determined well from Eq. (\ref{castaing2}). As shown in Fig. \ref{fig6} for {\it On1}, {\it On 3} and {\it Off} rather small $k$-values of $2.0$, $1.8$ and $3.3$ are obtained indicating that  resulting PDFs are heavy-tailed also for large scales. In Fig. \ref{fig8}a) and \ref{fig8}b) and Fig. \ref{fig8}d) the corresponding PDFs as well as fits according to Eq. (\ref{castaing2}) are presented. The fits agree quite well with measured distributions. For all three data sets the tails show a similar decay for large scales and are close to a straight line which corresponds to an exponential decay due to the semi-logarithmic presentation.\\
In contrast to this behaviour the PDFs of {\it On2} and the fits according to Eq. (\ref{castaing2}) -- shown in Fig. \ref{fig8}c) --  approach a Gaussian distribution for large scales which is in accordance with the large $k$-value of $8.7$ as it is found in Fig. \ref{fig6}. This large $k$-value mirrors the fact that {\it On2} was measured just over a short period of time (approximately $1\;hr$) with rather constant flow conditions. This means that as the flow becomes more and more stationary the superposition of different mean velocities becomes weaker and weaker. In the limit of a stationary velocity -- as realized for {\it Lab} and approximately for{\it On2}  -- isotropic turbulence is recovered and large scale intermittency disappears.\\

\noindent Note, the intention of using Eq. (\ref{castaing2}) is not to get the best possible fits of the measured PDFs but to model PDFs at specific locations (with a specific mean velocity distribution) {\it a priori}. 
The measured PDFs could also be fitted by stretched exponential distributions \cite{sornette} or by Eq. (\ref{castaing}) as it is done in Fig. \ref{fig3}b) for instance. These fits are based on {\it a posteriori} measurements however. Neither scaling models nor Eq. (\ref{castaing}) are able to explain a flatness larger than $3$ and correspondingly intermittent PDFs for large scales outside of the inertial range as it is done by Eq. (\ref{castaing2}). \\
This has interesting consequences for atmospheric velocities. For isotropic turbulence statistical properties of the flow can well be modelled. Thus knowing the composition of isotropic subsets it should be possible to model atmospheric velocity statistics as well.

\clearpage

\section{Conclusions}	\label{conc}

The presented analysis provides a good way to estimate and to explain robust and markedly intermittent PDFs of atmospheric increments. As desired the method recovers results of isotropic turbulence but is also able to take the large scale variations into account properly. A rigorous separation  of time scales -- which is very difficult -- is not necessary for our model.\\
Standard analysis applied to isotropic turbulence cannot be transferred to long atmospheric time series unambiguously as it was shown by means of scaling behaviour of structure functions and flatness. These methods are restricted to the inertial range where the change of shape can be more or less well reproduced and where at the large scale boundary of this range statistics become Gaussian.\\
It has been shown that the observed anomalous robust intermittency can be explained as a superposition of isotropic subsets in accordance with the concept of stable distributions. Intermittency on large scales is found to be the result of large scale mixing of isotropic turbulent subsets. \\ 
With our approach intermittent atmospheric PDFs can be approximated for any location as long as the mean velocity distribution is known. This is of importance for the construction of a wind turbine for instance. For constructing wind turbines only the turbulence intensity (relation between the average standard deviation and the mean velocity) is taken into account which is a very inadequate description.  The loads are determined by increments and their occurrence statistics, therefore a model that reproduces the right increment distributions for every location should be preferred.   

\subsection*{Acknowledgments} 
We acknowledge helpful discussions with {\it M. Siefert} and {\it B. Lange}. For data supply we further acknowledge {\it Ris{\o} National Laboratory} (offshore site funded by Energi E2/Seas), {\it Tauernwind Windkraftanlagen GmbH}, {\it Energiewerkstatt GmbH} and {\it J. Cleve} and {\it K. R. Sreenivasan}.


\begin{thebibliography}{5}

\bibitem{burton}
T. Burton, D. Sharpe, N. Jenkins and E. Bossanyi, {\it Wind Energy Handbook}, John Wiley \& Sons, 2001.

\bibitem{riso}
M. Nielsen, G.C. Larsen, J. Mann, S. Ott, K.S. Hansen and B.J. Pedersen: 'Wind Simulation for Extreme and Fatigue Loads', Riso-R-1437 (EN), 2003.

\bibitem{kantz}
M. Ragwitz \& H. Kantz: {\it Phys. Rev. Lett.} {\bf 87}, 254501, 2001.

\bibitem{us} 	
F. Boettcher, C. Renner, H.-P.Waldl and J. Peinke: 'On the statistics of wind gusts', {\it Boundary Layer Meteorology} 		{\bf 108}, 163-173, 2003.

\bibitem{bier} 	
W. Bierbooms, P.-W. Cheng: 'Stochastic gust model for design calculations of wind turbines', {\it Journal of Wind Engineering and Industrial Aerodynamics} {\bf 90}, 1237-1251, 2002.

\bibitem{vander} 	
I. Van der Hoven: {\it J. Meteorol.} {\bf 14}, 160-164, 1957.

\bibitem{lode1} 	
S. Lovejoy, D. Schertzer \& J.D. Stanway:  {\it Phys. Rev. Lett.} {\bf 86}, 5200-5203, 2001.

\bibitem{bush}
E. D. Egglestone \& R. N. Clark: {\it Wind Engineering} {\bf 24}, 49, 2000.

\bibitem{castaing} 
B. Castaing, Y. Gagne, E.J. Hopfinger: {\it Physica D} {\bf 46}, 177, 1990.

\bibitem{frisch}	
 U. Frisch: Turbulence. The Legacy of A. N. Kolmogorov, Cambridge University Press, 1995. 	

\bibitem{sornette}
 D. Sornette: Critical Phenomena in Natural Sciences, Springer, 2000.

\bibitem{adp} 	
J. Peinke, F. B\"ottcher, St. Barth: 'Anomalous statistics in turbulence, financial markets and other complex systems', {\it Ann. Phys.} {\bf 13}, No. 7-8, 450-460, 2004. 

\bibitem{hohlen} 	
H. Hohlen \& J. Liersch: ÔSynchrone Messkampagnen von Wind- und Windkraftanlagendaten am Standort FH Ostfriesland, EmdenÕ, {\it DEWI Magazin} {\bf 12}, 66Ð74, 1998. 

\bibitem{kleve} 
G. Stolovitzky, P. Kailasnath, K. R. Sreenivasan: 'Kolmogorov's Refined Similarity Hypotheses', {\it Phys. Rev. Lett.} {\bf 69}, 1178-1181, 1992.

\bibitem{dewi} 
H. Mellinghoff: 'Data Evaluation of a Wind Measurement equipped with Cup and Sonic Anemometers at Oberzeiring, Austria', {\it Report No.: DEWI-SO 0110-19}, 2001.

\bibitem{rebecca} 
B. Lange, R. Barthelmie, J. H{\o}jstrup: Description of R{\o}dsand field measurement, {\it Ris{\o}-R-1268(EN)}.

\bibitem{lueck} 
St. L\"uck: 'Skalenaufgel\"oste Experimente und statistische Analysen von turbulenten Nachlaufstr\"omungen', PhD thesis, Oldenburg, 2000.

\bibitem{christoph}
C. Renner: 'Markowanalysen stochastisch fluktuierender Zeitserien', PhD thesis, Oldenburg, 2002.
	
\bibitem{k41} 	
A. N. Kolmogorov: 'The Local Structure of Turbulence in an Incompressible Viscous Flow for Very High Reynolds 			Numbers', {\it Dokl. Acad. Nauk. SSSR} {\bf 30}, 301-305, 1941.

\bibitem{k62} 	
A. N. Kolmogorov: 'A refinement of previous hypotheses concerning the local structure of turbulence in a viscous incompressible fluid at high Reynolds number', {\it J. Fluid Mech.} {\bf 13}, 82, 1962.	

\bibitem{mu}
A. Arneodo et al.: 'Structure functions in turbulence, in various flow configurations, at Reynoldsnumber between 30 and 5000, using extended self-similarity', {\it Europhys. Lett.} {\bf 34} (6), 411, 1996.

\bibitem{she}
Z.-S. She \& E. Leveque: {\it Phys. Rev. Lett.} {\bf 72}, 336, 1994. 

\bibitem{lode}
F. Schmitt \& D. Schertzer, S. Lovejoy, Y. Brunet: 'Estimation of universal multifractal indices for atmospheric turbulent velocity fields', {\it Fractals} {bf 1} (3), 568-575, 1993.

\bibitem{yakhot} V. Yakhot, Phys. Rev. E {\bf 57}(2), 1737 (1998).

\bibitem{benzi} 	
R. Benzi, S. Ciliberto, R. Tripiccione, C. Baudet \& S. Succi: {\it Phys. Rev. E}, {\bf 48} (1), 29, 1993.

\bibitem{chabaudermann}
Chabaud, B., Naert, A., Peinke, J., Chilla, F., Castaing, B. and H\'ebral, B.: 
'A transition to developed turbulence',
{\it Phys. Rev. Lett.} {\bf 73}, 3227-3230, 1994.

\bibitem{beck}
C. Beck: 'Lagrangian acceleration statistics in turbulent flows', arXiv:cond-math/0212566v1, 2003.

\bibitem{gauss}
Ch. Renner, J. Peinke, R. Friedrich: 'Markov properties of small scale turbulence', {\it J. Fluid Mech.} {\bf 433}, 383, 2001. 

\bibitem{levy}
P. L\'evy: 'Theorie de l'Addition des Variables Al\'eatoires, Gauthier-Villards, Paris, 1937.

\bibitem{paul}	
 W. Paul, J. Baschnagel: Stochastic Processes -- From Physics to Finance, Springer, 1999. 

\bibitem{trevino}
G Trevi\~no and E. L. Andreas: 'Averaging intervals for spectral analysis of nonstationary turbulence', {\it Boundary- Layer Meteorol.} {\bf 95}, 231-247, 2000.















\end{thebibliography}
\end{document}